\documentstyle[aps,prl,preprint]{revtex} 
\begin{document}
\draft

\title{Extended parametric resonances in Nonlinear Schr\"odinger
  systems}

\author{Ju\'an J. Garc\'{\i}a-Ripoll, V\'{\i}ctor M.
  P\'erez-Garc\'{\i}a}

\address{Departamento de Matem\'aticas, Escuela T\'ecnica Superior de
  Ingenieros Industriales, \\
  Universidad de Castilla--La Mancha, 13071 Ciudad Real, Spain.}

\author{Pedro Torres}

\address{Departamento de Matem\'atica Aplicada, Facultad de Ciencias, \\
  Universidad de Granada, 18071 Granada, Spain.}

\date{\today}

\maketitle

\begin{abstract}
  We study an example of exact parametric resonance in a extended
  system ruled by nonlinear partial differential equations of
  nonlinear Schr\"odinger type. It is also conjectured how related
  models not exactly solvable should behave in the same way.  The
  results have applicability in recent experiments in Bose-Einstein
  condensation and to classical problems in Nonlinear Optics.
\end{abstract}


\narrowtext


Resonances are one of the recurrent themes of Physics. After reading
the current PACS sectioning scheme one finds that the word resonance
appears in fifty different items which is only a naive way to measure
one very important concept in Physics.

When speaking of resonances we use to refer to an ``anomalously large
response to a (maybe small) external perturbation''. The best known
model is the undamped harmonic oscillator driven by an external
force, which is textbook material. In this trivial system one easily
proves existence of unbounded linearly increasing terms in the
solution. A more complicated kind of resonant behavior are
parametric resonances, where a relevant parameters of the system is
modulated to achieve resonance. This phenomenon has been known since
the Middle Ages \cite{Sanmartin}, and its simplest examples are the
parametrically forced harmonic oscillator modeled by Hill's equation
$\ddot{x} + p(t) x = 0$ \cite{Hill} and a particular
version, $p(t) = 1 + \epsilon \cos \omega t$, known as Mathieu's
equation.

Since those are linear problems much can be said on their solutions.
For instance, Mathieu's equation can be studied by means of Floquet's
theory for periodic coefficient equations and resonance existence can
be proven analytically -even in the presence of linear dissipation-
for various parameter regions on the $(\epsilon, \omega)$
plane. However resonance phenomena are not restricted to linear
problems but also appear in nonlinear finite dimensional problems,
e.g. in many Hamiltonian chaotic systems where torus resonances are
the reason for the origin of chaos, in impact oscillators
and in nuclear magnetic and spin resonances, to cite only a few
examples.  There is finally the framework of systems modeled by
partial differential equations. Here some results are known in simple
linear cases, but in general the study of resonances in extended
(i.e. ruled by partial differential equations) nonlinear systems poses
many open questions.

Our aim in this letter is to prove that unbounded resonances are
possible in simple, experimentally realizable Hamiltonian wave
problems, specifically in the framework of nonlinear Schr\"odinger
(NLS) equations \cite{Vazquez}.  This is a very surprising result
since one is tempted to think (as it usually happens) that a resonant
perturbation can only excite a small number of modes of the infinite
present in an extended problem. Since the nonlinearities mix
different modes the amplification of one of them is stopped by energy
transfer to the others, specially in conservative systems where the
conservation laws control the number and amplitude of active modes.
Thus, it could be thought that once the energy is transfered to the
non-resonant modes and being their growth limited by nonlinear
constraints, the action of the perturbation should be controlled and
unbounded growth of the relevant quantities should be
inhibited. Indeed this is the common mechanism in many systems. We
will show that it is not the only possibility and truly resonant
behavior is possible even in simple physically relevant models.

{\em The model.-} Let us consider the following Nonlinear
Schr\"odinger equation in $n$ spatial dimensions.
\begin{equation}
\label{model3d}
  i \frac{\partial u}{\partial t} = - \frac{1}{2} \triangle u + |u|^2 u + V({\mathbf{r}},t) u,
\end{equation}
where $\triangle = \sum_{j=1}^n \partial^2/\partial x_j^2$.
This equation is the adimensional description of many different
phenomena \cite{Vazquez}. One of the hottest topics where it appears
is in the mean field model of nonuniform trapped Bose-Einstein
condensates (BEC) \cite{bose} where $V({\mathbf{r}}) =
\frac{1}{2}\sum_{j=1}^n \lambda_j(t) x_j^2$. Time dependent coefficients
reflect experiments in which the trap is perturbed to obtain the
response spectrum of the condensate \cite{expfreq} (For a theoretical
analysis of these experiments see Ref. \cite{Stringari}).  A more
detailed study of the condensate response to time dependent
perturbations was performed in \cite{Juanjo} and the existence of
strong resonances was proposed on the ground of approximate
variational methods and numerical simulations of the Eq.
(\ref{model3d}). Nonetheless, the nature of the resonance could not be
completely assured on the ground of approximate or numerical
techniques since both have limited applicability, specially for the
strong response phenomena considered here.

Let us first consider in this letter the two-dimensional case with
time-dependent parabolic potential
\begin{equation}
\label{model2d}
i \frac{\partial u}{\partial t} = -\frac{1}{2} \triangle u +
|u|^2 u + \frac{1}{2} \lambda(t)\left(x^2 + y^2 \right) u,
\end{equation}
where $\triangle = \partial^2/\partial x^2 + \partial^2/\partial y^2$.
Although Bose-Einstein condensation problems are usually three
dimensional, this model can be related to pancake type traps, and to
recent experiments with quasi two-dimensional condensates or
coherent atomic systems \cite{Gauck,Safonov}. The model
(\ref{model2d}) is also known in nonlinear Optics where it is
used to study the propagation of paraxial beams in fibers with a
(modulated) parabolic profile of the refraction index.

From the mathematical point of view, Eq. (\ref{model2d}) is a
parametrically forced NLS equation that has not been studied
rigorously before. Related systems are the externally driven damped
NLS in one dimension \cite{Kaup}, the AC-driven damped sine-Gordon
system \cite{Kaup2} and one type of non-self-adjoint parametrically
driven NLS \cite{Igor1,Igor2}. However those works concentrate on one
dimensional equations and mostly on soliton stability problems.

{\em Moment method.-} We will first study the radially symmetric
version of Eq.  (\ref{model2d}), searching solutions of the form
$\psi(r,\theta,t)=u(r,t) e^{im\theta}$, which includes both the
typical radially symmetric problem corresponding to $m=0$, and vortex
line solutions, with $m\neq 0$. The simplified equation for $u$ is
\begin{equation}
  \label{modelradial}
  i \frac{\partial u}{\partial t} =
  - \frac{1}{2r} \frac{\partial}{\partial r}\left( r\frac{\partial u}{\partial  r}\right)
  + \left(\frac{m^2}{2r} + |u|^2 + \frac{\lambda(t)}{2} r^2 \right)u,
\end{equation}
Eq. (\ref{modelradial}) is non-integrable and has no exact solutions
even in the constant $\lambda$ case. Its solutions are stationary
points of the action $S = \int_{t_0}^ {t_1}{\cal L}(t) $, where
\begin{eqnarray}
\label{lagrange}
{\cal L}(t) & = &
\frac{i}{2} \int\left(u\frac{\partial u^*}{\partial t}-u^*\frac{\partial u}{\partial t}\right) d^2x
+ \int \frac{\lambda(t)}{2}r^2|u|^2 d^2x \nonumber \\
& + & \frac{1}{2}\int \left(\left|\frac{\partial u^*}{\partial r} \right|^2
+ \frac{m^2}{r^2}|u|^2 + |u|^4 \right) d^2x.
\end{eqnarray}
Let us define the following integral quantities \cite{Momenta}
\begin{mathletters}
  \label{momentos}
  \begin{eqnarray}
    I_1(t) & = & \int  |u|^2 d^2x,  \\
    I_2(t) & = & \int  |u|^2 r^2 d^2x, \\
    I_3(t) & = & i \int  \left[u\frac{\partial u^*}{\partial r} -
      u^*\frac{\partial u}{\partial r}\right] r d^2x,\\
    I_4(t) & = & \frac{1}{2}\int \left(\left|\nabla u\right|^2
      + \frac{m^2}{r^2}|u|^2 + |u|^4 \right) d^2 x,
  \end{eqnarray}
\end{mathletters}
where $d^2x = 2\pi r dr$ because of the symmetry.  These magnitudes
are related physically to the norm (intensity or number of particles),
width, radial momentum and energy of the wave packet. In the Optical
interpretation of Eq. (\ref{modelradial}) these quantities are known
as moments and are used in (usually approximate) calculations related
to beam parameters evolution \cite{Momenta,momenta2}.  It is
remarkable that the $I_j$ satisfy simple, and most important, closed
evolution laws
\begin{mathletters}
\label{evo_momentos}
\begin{eqnarray}
\frac{dI_1}{dt} & = & 0, \\
\frac{dI_2}{dt} & = & I_3, \\
\frac{dI_3}{dt} & = & -2 \lambda(t) I_2 + 4 I_4, \\
\frac{dI_4}{dt} & = & -\frac{1}{2} \lambda(t) I_3.
 \end{eqnarray}
\end{mathletters}
The first equation comes from the phase invariance of Eq.
(\ref{modelradial}) under global phase transformations and corresponds
to the $L^2$-norm conservation (in BEC it is interpreted as the
particle number conservation in the mean field model).  The other
equations can also be obtained in connection with the invariance of
the action (\ref{lagrange}) under symmetry transformations.

{\em Singular Hill's equation.-} Eqs. (\ref{evo_momentos}) form a
linear non-autonomous system for the unknowns, $I_j, j=1,...,4$ that
has several positive invariants under time evolution, of which the
most important is
\begin{equation}
\label{conserved}
Q = 2 I_4 I_2 - I_3^2/4 > 0.
\end{equation}
With the help of this quantity, the system (\ref{evo_momentos}) can be
reduced to a single equation for the most relevant parameter $I_2(t)$,
which is
\begin{equation}
\label{nolineali}
\frac{d^2 I_2}{dt^2} - \frac{1}{2I_2} \left(\frac{dI_2}{dt}\right)^2 +
2 \lambda(t)I_2 = \frac{Q}{I_2}.
\end{equation}
If we were able to solve Eq. (\ref{nolineali}) then the use of Eqs.
(\ref{evo_momentos}) would allow us to track the evolution of the other
ones. We can do so by defining $X(t) = \sqrt{I_2}$, whose physical
meaning is the wavepacket width, and substituting it into
(\ref{nolineali}). This procedure gives us
\begin{equation}
\label{singular}
\ddot{X}+ \lambda(t) X = \frac{Q}{X^3}.
\end{equation}
The resulting equation is a singular (nonlinear) Hill equation. In
general it cannot be solved explicitly (see Ref. \cite{sing} for other
examples of these equations), however in this case we are fortunate that
the general solution of Eq. (\ref{singular}) can be obtained
\cite{Pinney} and is given by \begin{equation}
\label{solucion}
X(t) = \sqrt{u^2(t) + \frac{Q}{W^2}v^2(t)},
\end{equation}
where $u(t)$ and $v(t)$ are the two linearly independent solutions of
the equation
\begin{equation}
\label{Hill}
 \ddot{x} + \lambda(t) x = 0,
\end{equation}
which satisfy $u(t_0) = X(t_0), \dot{u}(t_0) = X'(t_0), v(t_0) = 0,
v'(t_0) \neq 0$, and $W$ is the Wronskian $W = u \dot{v} - \dot{u} v = \hbox{const.} \neq 0.$

The conclusion is that the evolution of the width of the wave packet,
which is given by (\ref{solucion}) is closely related to the solutions
of the Hill equation (\ref{Hill}). This is a very well studied problem
\cite{Hill} which is explicitly solvable only for particular choices
of $\lambda(t)$, but whose solutions are well characterized and many
of its properties are known.

A practical application of Eq. (\ref{solucion}) is that one may design
$\lambda(t)$ starting from the desired properties for the wave packet
width evolution. This can be done in BEC applications by controlling
the trapping potential and in Optics by the precise control of the
$z-$dependent refraction index of the waveguide.

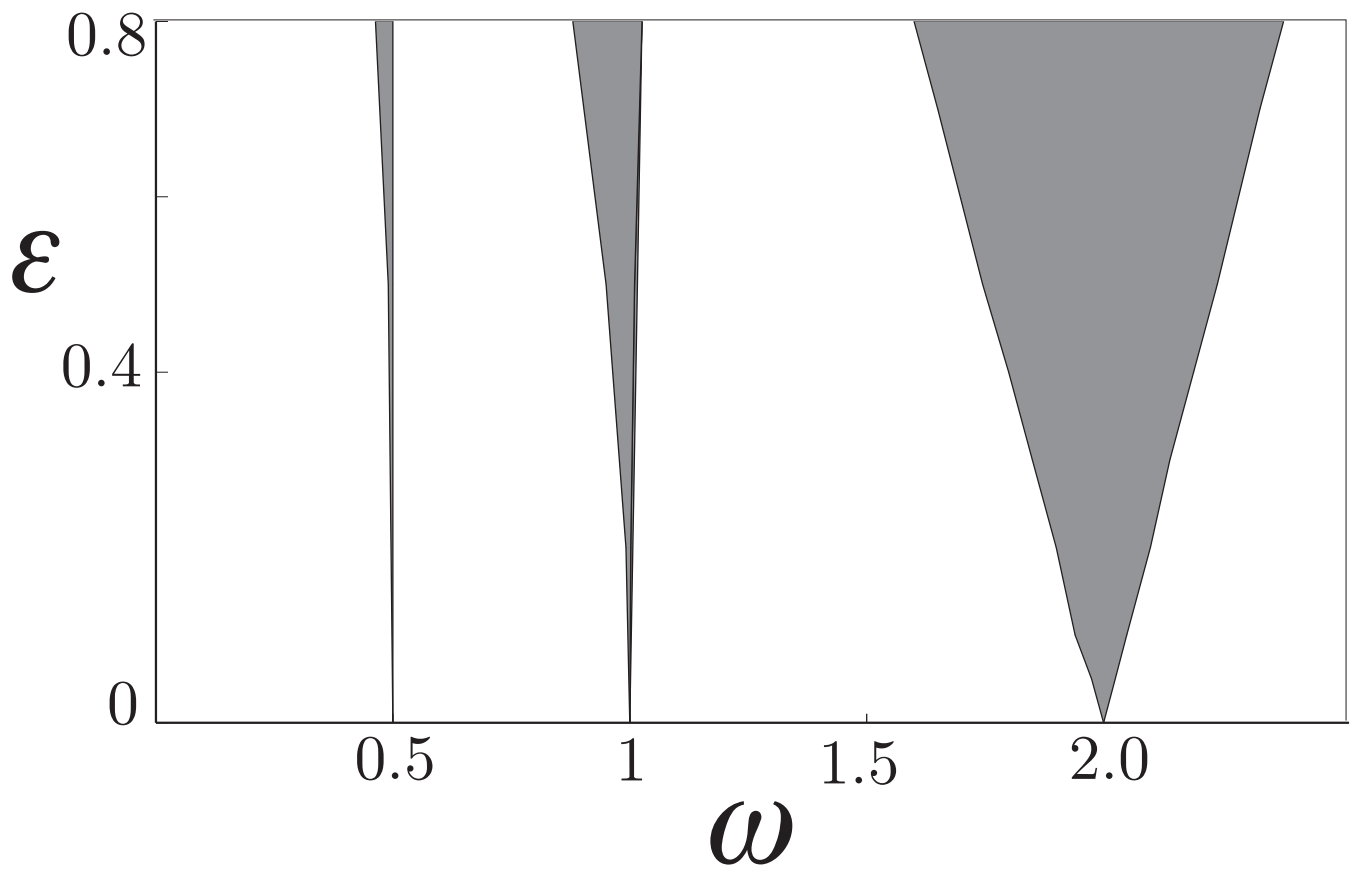
\begin{figure}
\caption{Stability diagram in the $\epsilon-\omega$ plane. The
instability regions corresponding to the first higher frequency bands
are shaded}
\label{one}
\end{figure}

If it is supposed that $\lambda(t)$ depends on a parameter
$\lambda(t)=1 + \tilde \lambda(t)$ with $\tilde \lambda(t)$ a periodic
function with zero mean value and peak value $\epsilon$ (not
necessarily small), then there is a complete theory describing intervals of
$\epsilon$ for which all solutions of Eq. (\ref{Hill}) are bounded
(stability intervals) and intervals for which all solutions are
unbounded (instability intervals). Both types of intervals are ordered
in a natural way.
Let us finally concentrate in the physically relevant case 
$\lambda(t) = 1 + \epsilon \cos \omega t$, and where $\epsilon$ needs
not to be a perturbative parameter. In that case the
parameter regions where exact resonances occur can be studied by
several means. First, for any fixed $\epsilon$, there exist
infinite ordered sequences $\{\omega_n\},\, \{\omega_n'\}$ tending to
zero such that Eq. (\ref{Hill}) is resonant if $\omega$ belongs to
$(\omega_n,\omega_n')$ for some $n$. Second, when $\omega$ is
fixed, resonances appear for $\epsilon$ big enough. Further, a
stability diagram can be drawn in the $\epsilon-\omega$ plane as shown
in Fig. \ref{one}.  The boundaries of these regions are the so-called
{\em characteristic curves}, which can not be explicitly derived but
whose existence can be proven analytically: if $D(\epsilon,\omega)$ is
the discriminant of the equation, characteristic curves are obtained
by solving the equations $D(\epsilon,\omega)=2$ and
$D(\epsilon,\omega)=-2$. In particular, instability regions start on
the frequencies $\omega=2,1,1/2,\ldots,2/n^2,\ldots$ \cite{Hill} [Fig.
\ref{one}].

The resonant behavior depends only on the mutual relation of the
parameters but not on the initial data. Since a resonance in Eq.
(\ref{Hill}) automatically implies unbounded behavior in the solution
[Eq. (\ref{solucion})] we obtain the result that for resonant
parameters all solutions have divergent width provided they start from
finite values of the moments (which is the generic case). With respect
to stability, it is deduced from Massera's Theorem that if
$(\epsilon,\omega)$ belongs to a stability region of Eq. (\ref{Hill}),
there exists a periodic solution of Eq.  (\ref{singular}) which is
Lyapunov stable.

We emphasize that our analysis of the cilindrically symmetric problem 
performed up to now is completely rigorous. Now we will turn to approximations
to discuss other related problems.

{\em Non-symmetric problems.-} Throughout this paper we have made use
of a special symmetry for the confining potential and for the solution
in order to make our theory exact.  When these constrains are removed
and we turn to a general two or three-dimensional model
(\ref{model3d}), a corresponding set of coupled Hill's equations may
still be obtained by some kind of approximation, which can be scaling
laws \cite{Castin} or a variational ansatz \cite{Juanjo}.

A new approach involves {\em only} imposing the wave-function to have
a quadratic dependence in the complex phase
$\psi=|\psi|\text{exp}\left({i\sum_{jk}\alpha_{jk}x_j x_k}\right)$,
$j,k=1,...,3$, which is a less restrictive ansatz than those used
previously. By introducing this trial function into (\ref{lagrange}),
building Lagrange's equations for the parameters, and using some
transformations, one obtains
\begin{equation}
\label{hill3d}
\ddot{X_i} + \lambda_i(t)X_i = \frac{1}{X_i^3} + \frac{Q}{X_iX_1X_2X_3}.
\end{equation}
Here $X_i, i=1,2,3$ are the three root mean square radius of the
solution for each spatial direction, i. e. the extensions of the
previous $X(t)$ to the three dimensional problem. In this case the
$\lambda_i(t)$ need not to be equal and in fact experimental problems
in BEC involve different $\lambda$ values because of asymmetries of the
traps.

Eqs. (\ref{hill3d}) are not integrable and form a six-dimensional
nonautonomous dynamical system for which few things can be said
analytically. Nevertheless, the numerical study of the approximate
model (\ref{hill3d}) exhibits an extended family of resonances which
is more or less the Cartesian product of those of (\ref{solucion}),
with minor displacements due to the coupling.  There exist also
parameter regions of chaotic and periodic solutions. Numerical
simulations of Eq. (\ref{model3d}) confirm the
predictions of the simple model (\ref{hill3d}).

{\em Nonlinear spectrum and resonances.-} Although resonant behavior
has been proven in the radially symmetric two dimensional case, one
could try to make a qualitative explanation for the reason that this
behavior is also present in non-symmetric problems. Let us look for
stationary solutions $u({\mathbf{r}},t) = \phi({\mathbf{r}}) e^{-i\mu
  t}$ for the stationary trap, $V=V({\mathbf{r}})$, and write the
nonlinear eigenvalue problem
\begin{equation}
\label{eigen3d}
  \mu \phi = - \frac{1}{2} \triangle \phi + |\phi|^2 \phi + V({\mathbf{r}}) \phi.
\end{equation}
Let us assume that we may use these solutions to expand, in a possibly
non-orthogonal and time dependent way, any solution of the
non-stationary problem (\ref{model3d}). By doing so one finds that the
energy absorption process in the time dependent potential is ruled by
the separation between the eigenvalues, $\mu_i -\mu_j$, of any two
different modes of (\ref{eigen3d}). If these differences are
distributed in a random way, the perturbation is not efficient and
does not lead to resonances, but to chaos. On the other hand, if the
differences may be approximated by multiples of a fixed set of
frequencies, then an appropriate parametric excitation will induce a
sustained process of energy gain (and width growth) such as the one we
have observed.

We have studied the spectrum $\{\mu\}$ for the case of an axially 
symmetric potential $V({\mathbf{r}}) = \lambda_r (x^2 + y^2) + 
\lambda_z z^2$ in three dimensions 
using a variant of a pseudospectral scheme using 
the harmonic oscillator basis as described in \cite{Vortices}. Our 
results show that the spectrum exhibits an ordered 
structure, with different directions of uniformity that may be excited 
by the parametric perturbation. We have checked the computations in one, two and
three spatial dimensions without assumptions of symmetry on the solution with similar
results despite the fact that the spectrum becomes more complex as dimensionality 
increases. Details of these calculations will be
published elsewhere.

{\em Losses effects.-} When losses are included in Eq.
(\ref{modelradial}), e.g. by the addition of a new term of the type $i
\sigma u$, it is not possible to obtain a set of closed equations for
the momenta and thus one cannot solve exactly the problem. However, as
discussed in the preceding comments one still may further restrict the
analysis to the parabolic phase approximation to obtain the modified
equation $\ddot{X} + \lambda(t) X = Q(t)/X^3$, where $Q(t)$ is a
decreasing function satisfying $Q \rightarrow 1$ as $t \rightarrow
\infty$.  At least up to the range of validity of the approximation
one obtains parametric resonances since the behavior of the singular
term does not affect essentially the resonance relation.  Of course,
at the same time one has a decrease in the norm of the solution since
now $dI_1/dt = - 2 \sigma I_1$.

In conclusion we have studied a parametrically perturbed nonlinear
extended system. By using the moment technique for the cilindrically symmetric 
problem we obtain a singular Hill equation which is reducible to a linear Hill equation and thus
existence of resonances can be shown analytically. For the physically
relevant case of a periodic perturbation it is shown that there
exist strong extended resonances for the relevant parameters of the
solution even when the solution is constrained by conservation
laws.  It is conjectured using the parabolic ansatz, analysis 
of nonlinear spectrum and numerical simulations that this behavior is also present
in non-radially symmetric problems or when losses are present.

V.M.P-G. has been partially supported by CICYT grants PB96-0534 and
PB95-0839. P. T. is supported by CICYT grant PB95-1203.


\begin{references}
  
\bibitem{Sanmartin}{J. Sanmart\'{\i}n, Am. Jour. Phys. {\bf 52}, 937
    (1984)}
  
\bibitem{Hill}{W. Magnus, S. Winkler, {\em Hill's equation}, Dover
    Publications, New York (1966); F.M. Arscott, {\em Periodic
      Differential Equations}, Pergamon Press, Oxford (1964)}
  
\bibitem{Vazquez}{{\em Nonlinear Klein-Gordon and Schr\"odinger
      systems: Theory and Applications}, L. V\'azquez, L. Streit, V.
    M. P\'erez-Garc\'{\i}a, Eds., World Scientific, Singapore (1996)}
  
\bibitem{bose}{M. H. Anderson, {\em et al.}, Science {\bf 269}, 198
    (1995); K. B. Davis, {\em et al.}, Phys. Rev. Lett. {\bf 75}, 3969
    (1995).}
  
\bibitem{expfreq}{ D. S. Jin, {\em et al.}, Phys. Rev. Lett.  {\bf 77}, 420 (1996); M.-O. Mewes, {\em et al.},
 {\em ibid.}, {\bf 77} 988 (1996); D. S. Jin, {\em et al.} {\em ibid}
    {\bf 78} 764 (1997)}
  
\bibitem{Stringari}{S. Stringari, Phys. Rev. Lett. {\bf 77}, 2360
    (1996); V.  M. P\'erez-Garc\'{\i}a, {\em et al.}, {\em ibid} {\bf
      77} 5320 (1996); V. M. P\'erez-Garc\'{\i}a, {\em et al.}, Phys.
    Rev. A {\bf 56} 1424 (1997)}
  
\bibitem{Juanjo}{J. J. Garc\'{\i}a-Ripoll, V. M. P\'erez-Garc\'{\i}a,
    Phys.  Rev. A {\bf 59} 2220 (1999)}
  
\bibitem{Gauck}{H. Gauck {\em et al.}, Phys. Rev. Lett. {\bf 81} 5298
    (1998)}
  
\bibitem{Safonov}{A. I. Safonov {\em et al.}, Phys. Rev. Lett. {\bf
      81} 4545 (1998)}
  
\bibitem{Kaup}{D. J. Kaup, A. C. Newell, Proc. R. Soc. London A {\bf
      361} 413 (1978); K. Nozaki, N. Bekki, Phys. Rev. Lett. {\bf 50}
    1226 (1983);  Physica D {\bf 21} 381 (1986)}
  
\bibitem{Kaup2}{D. J. Kaup, A. C. Newell, Phys. Rev. B {\bf 18} 5162
    (1978); A. R. Bishop {\em et al.} Physica D {\bf 40} 65 (1989); M.
    Taki {\em et al.}, Physica D {\bf 40} 65 (1989); G. Terrones {\em
      et al.} SIAM Jour. Appl. Math. {\bf 50} 791 (1990)}
  
\bibitem{Igor1}{V. E. Zaharov, V. S. L'vov, S. S. Starobinets, Sov.
    Phys.  Usp. {\bf 17} 896 (1975)}
  
\bibitem{Igor2}{I. Barashenkov, M. M. Bogdan, V. I. Korobov, Europhys.
    Lett. {\bf 15} 113 (1991); M. Bondila, I. Barashenkov, M. M.
    Bogdan, Physica D {\bf 87} 314 (1995)}
  
\bibitem{Momenta}{S. N. Vlasov, V. A. Petrischev, V. I. Talanov,
    Radiophys.  Quantum Electron. 14 (1971); P. A. Belanger, Opt.
    Lett.  {\bf 16} 196 (1991); M. A. Porras, Ph. D. Thesis,
    Universidad Complutense, Madrid (1993); M. A. Porras, J. Alda, E.
    Bernabeu, Appl. Opt. {\bf 32} 5885 (1993)}
  
\bibitem{momenta2}{V. M. P\'erez-Garc\'{\i}a, M. A. Porras, L.
    V\'azquez, Phys. Lett. A {\bf 202} (1995) 176}
  
\bibitem{sing}{J. Bevc, J. L. Palmer, C. S\"usskind, J. British Inst.
    Radio Engineers, {\bf 18} 696 (1958); T. Ding, Acta Sci. Natur.
    Univ. Pekin {\bf 11} 31 (1965); J. L. Reid, Proc. Am. Math. Soc.
    {\bf 27} 61 (1971); M. Zhang, J. Math. Anal. Appl. {\bf 203}
    (1996) 254; P. Torres, Math. Methods in Appl. Sci. (to appear).}
  
\bibitem{Castin}{Y. Castin and R. Dum, Phys. Rev. Lett. \textbf{77},
    5315 (1996).}
  
\bibitem{Vortices}{J. J. Garc\'{\i}a--Ripoll, V. M.
    P\'erez--Garc\'{\i}a, J. I. Cirac, Phys. Rev. A (submitted)}
  
\bibitem{Pinney}{E. Pinney, Proc. Am. Math. Soc. {\bf 1} 681 (1950)}

\end{references}
\end{document}